\documentclass[draftcls,12pt,onecolumn,oneside]{IEEEtran}
\usepackage{mathrsfs}
\usepackage{amsmath,amssymb,color,graphicx,epsfig,cite,psfrag,geometry}
\begin{document}
\title{Optimal Power Allocation for Fading Channels in
Cognitive Radio Networks: Ergodic Capacity and Outage Capacity}%
\author{Xin~Kang,~\IEEEmembership{Student
Member,~IEEE}, Ying-Chang~Liang,~\IEEEmembership{Senior
Member,~IEEE}, \\Arumugam~Nallanathan,~\IEEEmembership{Senior
Member,~IEEE}, \\Hari Krishna~Garg,~\IEEEmembership{Senior
Member,~IEEE}, and Rui~Zhang,~\IEEEmembership{Member,~IEEE}
\thanks{X. Kang and H. K. Garg are with the Department of Electrical $\&$ Computer
Engineering, National University of Singapore, 119260, Singapore
(Email: kangxin@nus.edu.sg, eleghk@nus.edu.sg).}
\thanks{Y.-C. Liang and R. Zhang are with Institute for Infocomm Research, 1 Fusionopolis Way,
$\sharp$21-01 Connexis, South Tower, Singapore 138632 (Email:
ycliang@i2r.a-star.edu.sg, rzhang@i2r.a-star.edu.sg).}
\thanks{A. Nallanathan is with the Division of Engineering, King's College London, London, United Kingdom
(Email: nallanathan@ieee.org).} }
\maketitle 

\begin{abstract}
A cognitive radio network (CRN) is formed by either allowing the
secondary users (SUs) in a secondary communication network (SCN) to
opportunistically operate in the frequency bands originally
allocated to a primary communication network (PCN) or by allowing
SCN to coexist with the primary users (PUs) in PCN as long as the
interference caused by SCN to each PU is properly regulated. In this
paper, we consider the latter case, known as spectrum sharing, and
study the optimal power allocation strategies to achieve the ergodic
capacity and the outage capacity of the SU fading channel under
different types of power constraints and fading channel models. In
particular, besides the interference power constraint at PU, the
transmit power constraint of SU is also considered. Since the
transmit power and the interference power can be limited either by a
peak or an average constraint, various combinations of power
constraints are studied. It is shown that there is a capacity gain
for SU under the average over the peak transmit/interference power
constraint. It is also shown that fading for the channel between SU
transmitter and PU receiver is usually a beneficial factor for
enhancing the SU channel capacities.
\end{abstract}



\section{Introduction}\label{intro}

Radio spectrum is a precious and limited resource for wireless
communication networks. With the emergence of new wireless
applications, the currently deployed spectrum is becoming
increasingly more crowded. Hence, how to accommodate more wireless
services within the limited spectrum becomes a challenging problem.
On the other hand, according to the report published by the Federal
Communication Commission (FCC), most of the allocated spectrum today
is under-utilized \cite{FCC}. This fact indicates that it is perhaps
the inefficient and inflexible spectrum allocation policy rather
than the physical shortage of spectrum that causes the spectrum
scarcity.

Cognitive radio (CR) \cite{Mitola} is a promising technology to deal
with the spectrum under-utilization problem caused by the current
inflexible spectrum allocation policy. In a cognitive radio network
(CRN), a secondary user (SU) in the secondary communication network
(SCN) is allowed to access the spectrum that is originally allocated
to the primary users (PUs) when the spectrum is not used by any PU.
This secondary spectrum usage method is called \emph{opportunistic
spectrum access} \cite{Haykin2005}. In this way, the spectrum
utilization efficiency can be greatly improved. However, to
precisely detect a vacant spectrum is not an easy
task\cite{LiangSensingTroughput}. Alternatively, CRN can also be
designed to allow simultaneous transmission of PUs and SUs. From
PU's perspective, SU is allowed to transmit as long as the
interference from SU does not degrade the quality of service (QoS)
of PU to an unacceptable level. From SU's perspective, SU should
control its transmit power properly in order to achieve a reasonably
high transmission rate without causing too much interference to PU.
This transmission strategy is termed as \emph{spectrum sharing}
\cite{Ghasemi}.

Traditionally, the capacity of fading channels is studied under
various transmit power constraints, and the corresponding optimal
and suboptimal power allocation policies are given in, e.g.,
\cite{Goldsmith}, \cite{Biglieri1998}, \cite{Liang2006}. Recently,
study on the channel capacity of SU link under spectrum sharing has
attracted a lot of attention. Specifically, SU channel capacity
under spectrum sharing was addressed by Gastpar in \cite{Gastpar},
where the capacities of different additive white Gaussian noise
(AWGN) channels are derived under a received power constraint. The
capacities derived in \cite{Gastpar} are shown to be quite similar
to those under a transmit power constraint. This is non-surprising
because the ratio of the received power to the transmit power is
fixed in an AWGN channel; thus, considering a received power
constraint is equivalent to considering a transmit power constraint.
However, in the presence of fading, the situation becomes quite
different. In \cite{Ghasemi}, the authors derived the optimal power
allocation strategy for a SU coexisting with a PU subject to an
interference power constraint at PU receiver, and evaluated the
ergodic capacity for SU channel for different fading channel models.
In \cite{Leilaglobalcom07}, the authors considered the outage
capacity under both the peak and the average interference power
constraints. It is noted that optimal design of SU transmission
strategy under interference-power constraints at PU receivers has
also been studied in \cite{Zhangrui} for multi-antenna CR
transmitters, and in \cite{ZhangLan} for multiple CR transmitters in
a multiple-access channel (MAC).

In this paper, we study the ergodic capacity, the delay-limited
capacity, and the outage capacity of SU block-fading (BF) channels
under spectrum sharing. For a BF channel
\cite{Ozarow1994,Caire1999}, the channel remains constant during
each transmission block, but possibly changes from one block to
another. For BF channels, the ergodic capacity is defined as the
maximum achievable rate averaged over all the fading blocks. Ergodic
capacity is a good performance limit indicator for delay-insensitive
services, when the codeword length can be sufficiently long to span
over all the fading blocks. However, for real-time applications, it
is more appropriate to consider the delay-limited capacity
introduced in \cite{Hanly}, which is defined as the maximum constant
transmission rate achievable over each of the fading blocks. For
certain severe fading scenarios, such as Rayleigh fading, however,
the delay-limited capacity could be zero. Thus, for such scenarios,
the outage capacity \cite{Ozarow1994,Caire1999}, which is defined as
the maximum constant rate that can be maintained over fading blocks
with a given outage probability, will be a good choice.

In this paper, we derive the optimal power allocation strategies for
SU to achieve aforementioned capacities. Besides the interference
power constraint to protect PU, we also consider the transmit power
constraint of SU transmitter. Since the transmit power and the
interference power can be limited either by a peak or an average
constraint, different combinations of power constraints are
considered. It is shown that there is a capacity gain for SU under
the average over the peak transmit/interference power constraint.
Furthermore, we provide closed-form solutions for the delay-limited
capacity and the outage probability under several typical channel
fading models, including Rayleigh fading, Nakagami fading, and
Log-normal fading. It is observed that fading for the channel
between SU transmitter and PU receiver can be a beneficial factor
for enhancing the SU channel capacities.

The rest of the paper is organized as follows. Section
\ref{system-model} describes the system model and presents various
transmit and interference power constraints. Then, the ergodic
capacity, the delay-limited capacity, and the outage capacity under
different combinations of peak/average transmit and interference
power constraints are studied in Section \ref{Ergodic-capacity},
Section \ref{section:delay-limited capacity}, and Section
\ref{sec:outage}, respectively. In Section \ref{NumericalResults},
the simulation results are presented and discussed. Finally, Section
\ref{conclusion} concludes the paper.

{\it Notation}: ${\rm E}[\cdot]$ denotes the statistical
expectation. $K$ denotes the constant $\log_{2}e$, where $e$ is the
base of natural logarithm. $\max(x,y)$ and $\min(x,y)$ denote the
maximum and the minimum element between $x$ and $y$, respectively.
$(\cdot)^+$ stands for $\max(0,\cdot)$. The symbol $\triangleq$
means ``defined as''.

\section{System Model and Power Constraints}\label{system-model}
\subsection{System model}
\begin{figure}[t]
        \centering
        \includegraphics*[width=12cm]{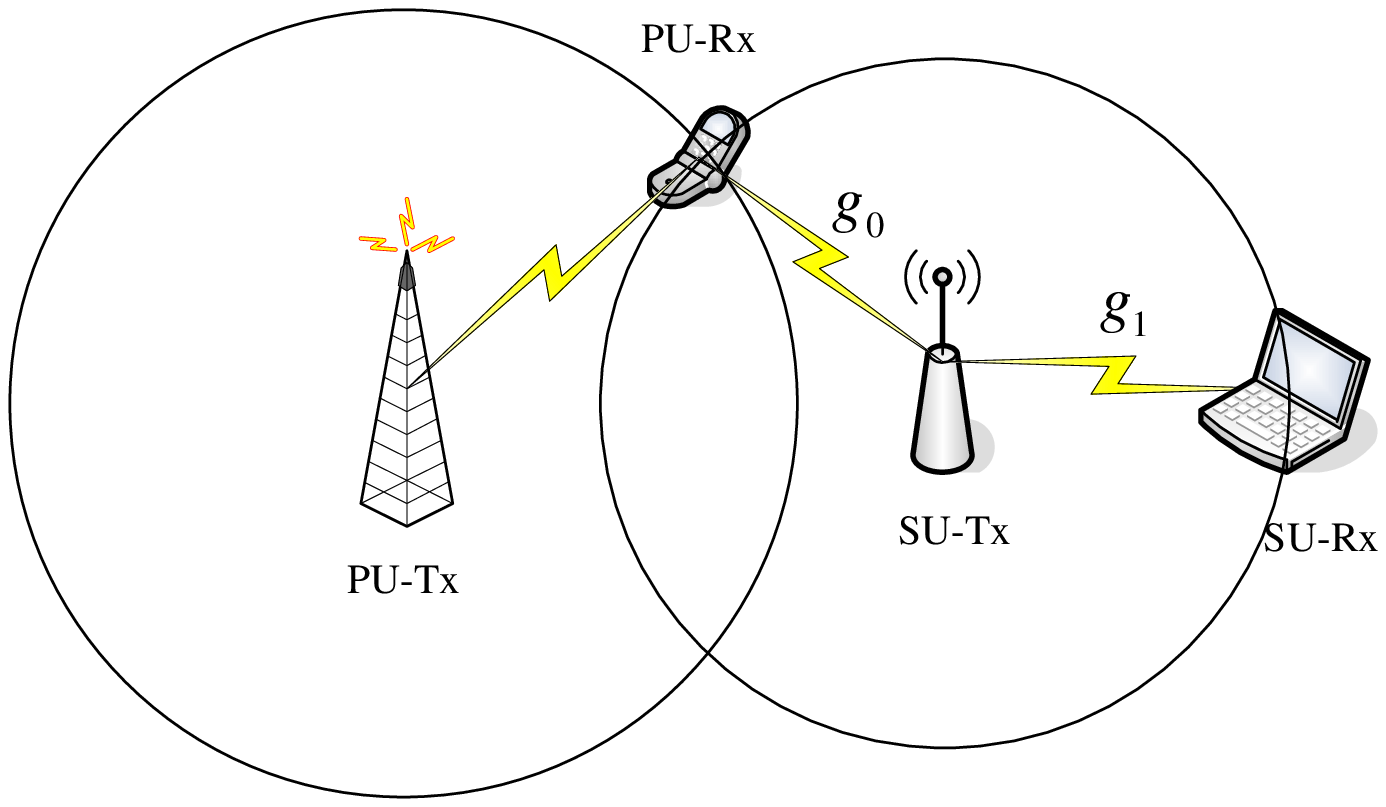}
        \caption{System model for spectrum sharing in cognitive radio networks.}
        \label{model}
\end{figure}

As illustrated in Fig. \ref{model}, we consider a spectrum sharing
network with one PU and one SU. The link between SU transmitter
(SU-Tx) and PU receiver (PU-Rx) is assumed to be a flat fading
channel with instantaneous channel power gain $g_0$ and the AWGN
$n_0$. SU channel between SU-Tx and SU receiver (SU-Rx) is also a
flat fading channel characterized by instantaneous channel power
gain $g_1$ and the AWGN $n_1$. The noises $n_0$ and $n_1$ are
assumed to be independent random variables with the distribution
${\mathcal C}{\mathcal N}(0,N_0)$ (circularly symmetric complex
Gaussian variable with mean zero and variance $N_0$). The channel
power gains, $g_0$ and $g_1$, are assumed to be ergodic and
stationary with probability density function (PDF) $f_0(g_0)$, and
$f_1(g_1)$, respectively. Perfect channel state information (CSI) on
$g_0$ and $g_1$ is assumed to be available at SU-Tx. Furthermore, it
is assumed that the interference from PU-Tx to SU-Rx can be ignored
or considered in the AWGN at SU-Rx.

\subsection{Power constraints}
Previous study on the fading channel capacity usually assumes two
types of power constraints at the transmitter: peak transmit power
constraint and average transmit power constraint, either
individually \cite{Caire1999} or simultaneously
\cite{Khojastepour2004b}. The peak power limitation may be due to
the nonlinearity of power amplifiers in practice, while the average
power is restricted below a certain level to keep the long-term
power budget. In this paper, we denote the instantaneous transmit
power at SU-Tx for the channel gain pair $(g_0,g_1)$ as
$P(g_0,g_1)$, and obviously it follows
\begin{align}\label{transmit-pk-largero}
P(g_0,g_1)\ge 0, \forall (g_0,g_1).
\end{align}

Let $P_{pk}$ be the peak transmit power limit and $P_{av}$ be the
average transmit power limit. The peak transmit power constraint can
then be represented by
\begin{align}\label{con-transmit-pk}
P(g_0,g_1)\le P_{pk}, \forall (g_0,g_1),
\end{align}
and the average transmit power constraint can be represented by
\begin{align}\label{con-transmit-av}
{\rm E}[P(g_0,g_1)]\le P_{av}.
\end{align}

On the other hand, motivated by the interference temperature concept
in \cite{Haykin2005}, researchers have investigated SU channel
capacities with received power constraints. If PU provides
delay-insensitive services, an average received power constraint can
be used to guarantee a long-term QoS of PU. Let $Q_{av}$ denote the
average received power limit at PU-Rx. The average interference
power constraint can then be written as
\begin{align}\label{con-interference-av}
{\rm E}[g_0P(g_0,g_1)]\le Q_{av}.
\end{align}
If the service provided by PU has an instantaneous QoS requirement,
the peak interference power constraint may be more appropriate. Let
$Q_{pk}$ denote the peak received power at the PU-Rx. The peak
interference power constraint can then be written as
\begin{align}\label{con-interference-pk}
g_0P(g_0,g_1)\le Q_{pk}, \forall (g_0,g_1).
\end{align}

For the purpose of exposition, we combine the transmit power
constraint with the interference power constraint, and obtain the
following four sets of power constraints:
\begin{align}
&\mathscr{F}_{1}\triangleq\{P(g_0,g_1): \eqref{transmit-pk-largero},\eqref{con-transmit-pk}, \eqref{con-interference-pk}\},\\
&\mathscr{F}_{2}\triangleq\{P(g_0,g_1): \eqref{transmit-pk-largero},\eqref{con-transmit-pk}, \eqref{con-interference-av} \},\\
&\mathscr{F}_{3}\triangleq\{P(g_0,g_1): \eqref{transmit-pk-largero},\eqref{con-transmit-av}, \eqref{con-interference-pk}\},\\
&\mathscr{F}_{4}\triangleq\{P(g_0,g_1):
\eqref{transmit-pk-largero},\eqref{con-transmit-av},
\eqref{con-interference-av} \}.
\end{align}

\section{Ergodic Capacity}\label{Ergodic-capacity}
For BF channels, ergodic capacity is defined as the maximum
achievable rate averaged over all the fading blocks. Using a similar
approach as in \cite{Goldsmith}, the ergodic capacity of the
secondary link can be obtained by solving the following optimization
problem,
\begin{eqnarray}\label{ecase1}
\max_{P(g_0,g_1)\in\mathscr{F}}{{\rm
E}\left[\log_2{\left(1+\frac{g_1P(g_0,g_1)}{N_0}\right)}\right]},
\end{eqnarray}
where $\mathscr{F} \in
\left\{\mathscr{F}_{1},\mathscr{F}_{2},\mathscr{F}_{3},\mathscr{F}_{4}\right\}$,
and the expectation is taken over $(g_0,g_1)$. In what follows, we
will study (\ref{ecase1}) under $\mathscr{F}_{1}$,
$\mathscr{F}_{2}$, $\mathscr{F}_{3}$, and $\mathscr{F}_{4}$,
respectively.

\subsection{Peak transmit power
constraint and peak interference power constraint}\label{cabothpeak}
In this case, $\mathscr{F}$ in \eqref{ecase1} becomes
$\mathscr{F}_{1}$. The two constraints in $\mathscr{F}_{1}$ can be
combined as $P(g_0,g_1)\le \min\{P_{pk}, \ \frac{Q_{pk}}{g_0}\}$.
Therefore, the capacity is maximized by transmitting at the maximum
instantaneous power expressed as
\begin{eqnarray}\label{pcase1}
P(g_0,g_1)=\left\{\begin{array}{lr}
P_{pk},&g_0 \le \frac{Q_{pk}}{P_{pk}}\\
\frac{Q_{pk}}{g_0},& \mbox{otherwise}
\end{array}
\right..
\end{eqnarray}


From (\ref{pcase1}), it is observed that, when $g_0$ is less than a
given threshold, SU-Tx can transmit at its maximum power, $P_{pk}$,
which satisfies the interference power constraint at PU-Rx. This
indicates that sufficiently severe fading of the channel between
SU-Tx and PU-Rx is good from both viewpoints of protecting PU-Rx and
maximizing SU throughput. However, when $g_0$ becomes larger than
this threshold, SU-Tx transmits with decreasing power values that
are inversely proportional to $g_0$.

\subsection{Peak transmit power constraint and average interference
power constraint}\label{captai}

In this case, $\mathscr{F}$ in \eqref{ecase1} becomes
$\mathscr{F}_{2}$. The optimal power allocation is given by the
following theorem.

\emph{Theorem 1:} The optimal solution of (\ref{ecase1}) subject to
the power constraints given in $\mathscr{F}_{2}$ is
\begin{eqnarray}\label{pcase2}
P(g_0,g_1)=\left\{\begin{array}{ll} 0,&g_0\ge \frac{Kg_1}{\lambda N_0}\\
\frac{K}{\lambda g_0}-\frac{N_0}{g_1},&\frac{Kg_1}{\lambda N_0}>g_0>\frac{K}{\lambda(P_{pk}+\frac{N_0}{g_1})}\\
P_{pk},&g_0\le \frac{K}{\lambda(P_{pk}+\frac{N_0}{g_1})}
\end{array}
\right.,
\end{eqnarray}
where $\lambda$ is the nonnegative dual variable associated with
\eqref{con-interference-av} in $\mathscr{F}_{2}$. If
\eqref{con-interference-av} in $\mathscr{F}_{2}$ is satisfied with
strict inequality, $\lambda$ must be zero. Otherwise, $\lambda$ can
be obtained by substituting (\ref{pcase2}) into the constraint ${\rm
E}[g_0P(g_0,g_1)]=Q_{av}$.

\begin{proof}
See Appendix A.
\end{proof}


As can be seen from \eqref{pcase2}, if $P_{pk}$ is sufficiently
large, the power allocation scheme reduces to that in
\cite{Ghasemi}, where the ergodic capacity of fading channels is
studied under the interference power constraint only. It is also
noticed that the power allocation scheme given by \eqref{pcase2} has
the same structure as that in \cite{Khojastepour2004b}, where the
ergodic capacity of fading channels is studied under both peak and
average transmit power constraints. The main difference is that the
power allocation scheme given by \eqref{pcase2} is not only related
to SU channel but also related to the channel between SU-Tx and
PU-Rx.

\subsection{Average transmit power constraint and peak interference
power constraint}\label{caatpi}

In this case, $\mathscr{F}$ in \eqref{ecase1} becomes
$\mathscr{F}_{3}$. The optimal power allocation of this problem is
given by the following theorem.

\emph{Theorem 2:} The optimal solution of (\ref{ecase1}) subject to
the constraints given in $\mathscr{F}_{3}$ is
\begin{eqnarray}\label{fpcase3}
P(g_0,g_1)=\left\{\begin{array}{ll}
0,&g_1\le \frac{\lambda N_0}{K}\\
\frac{K}{\lambda}-\frac{N_0}{g_1},&g_1> \frac{\lambda N_0}{K},g_0<\frac{Q_{pk}}{(\frac{K}{\lambda}-\frac{N_0}{g_1})}\\
\frac{Q_{pk}}{g_0}, &g_1> \frac{\lambda N_0}{K},g_0\ge
\frac{Q_{pk}}{(\frac{K}{\lambda}-\frac{N_0}{g_1})}
\end{array}
\right.,
\end{eqnarray}
where $\lambda$ is the nonnegative dual variable associated with
\eqref{con-transmit-av} in $\mathscr{F}_{3}$. If
\eqref{con-transmit-av} in $\mathscr{F}_{3}$ is satisfied with
strict inequality, $\lambda$ must be zero. Otherwise, $\lambda$ can
be obtained by substituting (\ref{fpcase3}) into the constraint
${\rm E}[P(g_0,g_1)]=P_{av}$.

Theorem 2 can be proved similarly as Theorem 1, we thus omit the
details here for brevity.


From (\ref{fpcase3}), it is seen that, when the channel between
SU-Tx and PU-Rx experiences sufficiently severe fading or $Q_{pk}$
is sufficiently large, the power allocation reduces to the
conventional water-filling solution \cite{Goldsmith}. It is also
observed that the power allocation given in \eqref{fpcase3} is
capped by $\frac{Q_{pk}}{g_0}$, and this cap increases with
decreasing $g_0$. This indicates that fading for the channel between
SU-Tx and PU-Rx enables SU-Tx to transmit more powers under the same
value of $Q_{pk}$.

\subsection{Average transmit power constraint and average interference power constraint}\label{caatai}
In this case, $\mathscr{F}$ in \eqref{ecase1} becomes
$\mathscr{F}_{4}$. The optimal solution for this problem can be
obtained by applying similar techniques as for Theorem 1, which can
be expressed as
\begin{eqnarray}\label{Pcase4}
P(g_0,g_1)=\left(\frac{K}{\lambda+\mu g_0}-\frac{N_0}{g_1}\right)^+,
\end{eqnarray}
where $\lambda$ and $\mu$  are the nonnegative dual variables
associated with \eqref{con-transmit-av} and
\eqref{con-interference-av} in $\mathscr{F}_{4}$, respectively. If
\eqref{con-transmit-av} or \eqref{con-interference-av} in
$\mathscr{F}_{4}$ is satisfied with strict inequality, $\lambda$ or
$\mu$ must be zero correspondingly. Otherwise, $\lambda$ and $\mu$
can be jointly determined by substituting (\ref{Pcase4}) into the
constraints ${\rm E}[P(g_0,g_1)]=P_{av}$ and ${\rm E}[g_0
P(g_0,g_1)]=Q_{av}$.

\section{Delay-limited Capacity}\label{section:delay-limited
capacity} For BF channels, delay-limited capacity \cite{Hanly} is
defined as the maximum constant transmission rate achievable over
each of the fading blocks. This is a good performance limit
indicator for delay-sensitive services, which may require a constant
rate transmission over all the fading blocks. Thus, the objective is
to maximize such constant rate by adapting the transmit power of
SU-Tx. At the same time, due to the coexistence with PU, the
received interference power at the PU-Rx should not exceed the given
threshold. In this section, the delay-limited capacity is studied
under $\mathscr{F}_{4}$ only. This is due to the fact that
delay-limited capacity can be shown to be zero under the other three
combinations of power constraints for realistic fading channel
models. Therefore, the delay-limited capacity can be obtained by
solving the following problem:
\begin{align}
\max_{P(g_0,g_1)\in \mathscr{F}_{4}}&\log_{2}\left(1+\gamma\right),\\
\mbox{s. t.}~~~&\frac{g_1P(g_0,g_1)}{N_0}=\gamma,~
\forall(g_0,g_1).\label{delay:gamma}
\end{align}
where $\gamma$ is the constant received signal-to-noise ratio (SNR)
at SU-Rx for all pairs of $(g_0,g_1)$.

Obviously, the delay-limited capacity is achieved when $\gamma$
takes its maximum value. Therefore, the above problem is equivalent
to finding the maximum value of $\gamma$ under the power constraints
in $\mathscr{F}_{4}$. From \eqref{delay:gamma}, we have
$P(g_0,g_1)=\frac{\gamma N_0}{g_1}$. Substituting this into the
power constraints given in $\mathscr{F}_{4}$ yields $\gamma \le
\frac{P_{av}}{N_0 {\rm E}\left\{\frac{1}{g_1}\right\}}$ and $\gamma
\le \frac{Q_{av}}{N_0 {\rm E}\left\{\frac{g_0}{g_1}\right\}}$,
$\forall(g_0,g_1)$. Therefore, $ \gamma_{max}=\min
\left\{\frac{P_{av}}{N_0 {\rm E}\left\{\frac{1}{g_1}\right\}},
\frac{Q_{av}}{N_0 {\rm E}\left\{\frac{g_0}{g_1}\right\}}\right\}. $
The delay-limited capacity is thus given by
\begin{align} \label{delaycapacity}
C_{d}\kern-1mm=\kern-1mm\min\left\{\log_{2}\left(1\kern-1mm+\kern-1mm\frac{P_{av}}{N_0
{\rm E}\left\{\frac{1}{g_1}\right\}}\right),
\log_{2}\left(1\kern-1mm+\kern-1mm\frac{Q_{av}}{N_0 {\rm
E}\left\{\frac{g_0}{g_1}\right\}}\right) \kern-1mm\right\}.
\end{align}

By setting $Q_{av}=+\infty$ in \eqref{delaycapacity}, it is easy to
obtain the delay-limited capacity for the conventional fading
channels \cite{Caire1999}. Similarly, by setting $P_{av}=+\infty$,
the delay-limited capacity under the interference power constraint
only is obtained.

In the following, the delay-limited capacity is evaluated under
different fading channel models.

\subsection{Rayleigh fading}

For Rayleigh fading, the channel power gains $g_0$ and $g_1$ are
exponentially distributed. Assume $g_0$ and $g_1$ are unit-mean and
mutually independent. Then, ${\rm E}\left[\frac{1}{g_1}\right]$ can
be evaluated equal to $+\infty$. Furthermore, the $\rm PDF$ of
$\frac{g_0}{g_1}$ is expressed as  \cite{Ghasemi}
\begin{align}\label{pdf-rayleigh}
f_{\frac{g_0}{g_1}}(x)=\frac{1}{(x+1)^{2}}, ~x\ge 0.\end{align}
Hence, ${\rm E}\left[\frac{g_0 }{g_1}\right]$ can be shown to be
$+\infty$. Therefore, from \eqref{delaycapacity}, the delay-limited
capacity is zero for Rayleigh fading channels.

\subsection{Nakagami fading}
Another widely used channel model is Nakagami-$m$ fading. For a
unit-mean Nakagami fading channel, the distribution of channel power
gain follows the Gamma distribution, which is expressed as
\begin{align}f_g(x)=\frac{m^m x^{(m-1)}}{\Gamma(m)}e^{-mx},~x\ge 0,\end{align} where
$\Gamma(\cdot)$ is the Gamma function defined as
$\Gamma(x)=\int_0^\infty t^{(x-1)}e^{-t}dt$, and $m$ $(m\ge1)$ is
the ratio of the line-of-sight (LOS) signal power to that of the
multi-path component. Then, by \cite{Tablesofintegral}, ${\rm
E}\left[\frac{1}{g_1}\right]$ is evaluated to be $1$.  If $g_0$ and
$g_1$ are independent and have the same parameter $m$, the PDF of
$\frac{g_0}{g_1}$ is  \cite{Nakagami}
\begin{align}\label{pdf-nakagami}
f_{\frac{g_0}{g_1}}(x)=\frac{x^{m-1}}{\mathcal {B}(m,m)(x+1)^{2m}},
~x\ge 0,\end{align} where $\mathcal {B}(a,b)$ is the Beta function
defined as $\mathcal
{B}(a,b)=\frac{\Gamma(a)\Gamma(b)}{\Gamma(a+b)}$. Then ${\rm
E}\left[\frac{g_0}{g_1}\right]$ can be evaluated equal to
$\frac{m}{m-1}$. Hence, the delay-limited capacity in
\eqref{delaycapacity} is obtained as
\begin{align}\label{eq:10}
C_{d}\kern-1mm=\kern-1mm\min\left\{\log_{2}\left(1\kern-1mm+\kern-1mm\frac{P_{av}}{N_0
}\right), \log_{2}\left(1\kern-1mm+\kern-1mm\frac{Q_{av}}{N_0
\frac{m}{m-1}}\right) \kern-1mm\right\}.
\end{align}

By setting $P_{av}=+\infty$, the delay-limited capacity under the
interference power constraint only is obtained as $ C_{d}=
\log_{2}\left(1+\frac{Q_{av}}{N_0 \frac{m}{m-1}}\right)$.
Furthermore, it is seen from \eqref{eq:10} that the delay-limited
capacity is determined by only the interference power constraint
when $P_{av}\ge \frac{m-1}{m}Q_{av}$.

\subsection{Log-normal shadowing}
In the log-normal fading environment, the channel power gain is
modeled by a log-normal random variable (r.v.) $e^X$ where $X$ is a
zero-mean Gaussian r.v. with variance $\sigma^2$. In this case, we
model the channel by letting $g_0=e^{X_0}$ and $g_1=e^{X_1}$, where
$X_0$ and $X_1$ are independently distributed with  mean zero and
variance $\sigma^2$. Under the above assumptions, $g_0/g_1=e^Y$ is
also log-normally distributed with $Y=X_0-X_1$ being Gaussian
distributed with mean zero and variance $2\sigma^2$
\cite{RandomProcess}. In this case, ${\rm E}[\frac{1}{g_1}]$ and
${\rm E}[\frac{g_0}{g_1}]$ are evaluated to be
$e^{\frac{\sigma^2}{2}}$ and $e^{\sigma^2}$, respectively. Hence,
the delay-limited capacity in \eqref{delaycapacity} is given by
\begin{align}\label{eq:11}
C_{d}\kern-1mm=\kern-1mm\min\left\{\log_{2}\left(1\kern-1mm+\kern-1mm\frac{P_{av}}{N_0
e^{\frac{\sigma^2}{2}}}\right),
\log_{2}\left(1\kern-1mm+\kern-1mm\frac{Q_{av}}{N_0
e^{\sigma^2}}\right) \kern-1mm\right\}.
\end{align}

By setting $P_{av}=+\infty$, the delay-limited capacity under the
interference power constraint only is obtained as $ C_{d}=
\log_{2}\left(1+\frac{Q_{av}}{N_0 e^{\sigma^2}}\right)$.
Furthermore, it is seen from \eqref{eq:11} that the delay-limited
capacity will not be affected by the transmit power constraint when
$P_{av}\ge e^{-\frac{\sigma^2}{2}}Q_{av}$.

\section{Outage Capacity}\label{sec:outage}

For BF channels, outage capacity is defined as the maximum rate that
can be maintained over the fading blocks with a given outage
probability. Mathematically, this problem is defined as finding the
optimal power allocation to achieve the maximum rate for a given
outage probability,
which is equivalent to minimizing the outage probability for a given
transmission rate (outage capacity) $r_0$, expressed as
\begin{align}\label{outage-probability}
\min_{P\left(g_0,g_1\right)\in\mathscr{F}} \quad
Pr\left\{\log_{2}\left(1+\frac{g_1
P\left(g_0,g_1\right)}{N_0}\right)< r_0\right\},
\end{align}
where $Pr\left\{\cdot\right\}$ denotes the probability.

In the following, we will study the problem
\eqref{outage-probability} under $\mathscr{F}_{1}$,
$\mathscr{F}_{2}$, $\mathscr{F}_{3}$, and $\mathscr{F}_{4}$,
respectively.


\subsection{Peak transmit power
constraint and peak interference power constraint}

In this case, $\mathscr{F}$ in \eqref{outage-probability} becomes
$\mathscr{F}_{1}$. The optimal solution of this problem can be
easily obtained as
\begin{align}\label{outage-power-pp}
P(g_0,g_1)\kern-1mm=\kern-1mm\left\{\kern-1mm\begin{array}{cc}
           \frac{N_0\left(2^{r_0}-1\right)}{g_1},&g_1\ge \frac{N_0\left(2^{r_0}-1\right)}{P_{pk}},g_0\le\frac{g_1Q_{pk}}{ N_0 (2^{r_0}-1)}\\
           0, &\mbox{otherwise}
         \end{array}
\right. .
\end{align}

Substituting \eqref{outage-power-pp} into
\eqref{outage-probability}, we get
\begin{align}
\mathscr{P}_{out}\kern-1mm=\kern-1mm1\kern-1mm-\kern-1mm\int_{\frac{
N_0(2^{r_0}-1)}{P_{pk}}}^{+\infty}\int_{0}^{\frac{g_1Q_{pk}}{N_0\left(2^{r_0}-1\right)}}
f_0(g_0)f_1(g_1)dg_0dg_1.
\end{align}

It is seen that \eqref{outage-power-pp} has the similar structure as
the truncated channel inversion \cite{Goldsmith} for the convenional
fading channel. The difference between these two methods lies in
that the condition in \eqref{outage-power-pp} for channel inversion
is determined by both $g_0$ and $g_1$, while that in
\cite{Goldsmith} is by $g_1$ only. Therefore, we refer to this power
allocation strategy as
\emph{two-dimensional-truncated-channel-inversion (2D-TCI)} over
$g_0$ and $g_1$.

\subsection{Peak transmit power constraint and average interference
power constraint}

In this case, $\mathscr{F}$ in \eqref{outage-probability} becomes
$\mathscr{F}_{2}$. The optimal solution of this problem is given by
the following theorem.

\emph{Theorem 3:} The optimal solution of \eqref{outage-probability}
subject to the power constraints given in $\mathscr{F}_{2}$ is
\begin{align}\label{outage-power-pa}
P(g_0,g_1)\kern-1mm=\kern-1mm\left\{\kern-1.5mm\begin{array}{cc}
           \frac{N_0\left(2^{r_0}-1\right)}{g_1},&g_1\ge \frac{N_0\left(2^{r_0}-1\right)}{P_{pk}},g_0 <\frac{g_1}{\lambda N_0 (2^{r_0}-1)}\\
           0, &\mbox{otherwise}
         \end{array}
\right.\kern-1mm,
\end{align}
and the corresponding minimum outage probability is given by
\begin{align}\label{26}
\mathscr{P}_{out}\kern-1mm=\kern-1mm1\kern-1mm-\kern-1mm\int_{\frac{N_0\left(2^{r_0}-1\right)}{P_{pk}}}^{+\infty}\int_0^{\frac{g_1}{\lambda
N_0 (2^{r_0}-1)}} f_0(g_0)f_1(g_1)dg_0dg_1,
\end{align} where $\lambda$ is the nonnegative dual variable associated
with \eqref{con-interference-av} in $\mathscr{F}_{2}$. If
\eqref{con-interference-av} in $\mathscr{F}_{2}$ is satisfied with
strict inequality, $\lambda$ must be zero. Otherwise, $\lambda$ can
be obtained by substituting \eqref{outage-power-pa} into the
constraint ${\rm E}\left[g_0 P(g_0,g_1)\right]=Q_{av}$.

\begin{proof} See Appendix B.\end{proof}

It is seen that \eqref{outage-power-pa} has the same structure as
that in \eqref{outage-power-pp}. Therefore, the optimal power
control policy obtained in \eqref{outage-power-pa} is also 2D-TCI.

\subsection{Average transmit power constraint and peak interference
power constraint}

In this case, $\mathscr{F}$ in \eqref{outage-probability} becomes
$\mathscr{F}_{3}$. The optimal solution of this problem is given by
the following theorem.

\emph{Theorem 4:} The optimal solution of \eqref{outage-probability}
subject to the power constraints given in $\mathscr{F}_{3}$ is
\begin{align}\label{outage-power-ap}
P(g_0,g_1)\kern-1mm=\kern-1mm\left\{\kern-2mm\begin{array}{cc}
           \frac{N_0\left(2^{r_0}-1\right)}{g_1},&g_1> \lambda N_0(2^{r_0}\kern-1mm-\kern-1mm1),g_0\le\frac{g_1Q_{pk}}{N_0\left(2^{r_0}-1\right)}\\
           0, &\mbox{otherwise}
         \end{array}
\right.\kern-2mm,
\end{align}
and the corresponding minimum outage probability is given by
\begin{align}
\mathscr{P}_{out}\kern-1mm=\kern-1mm1\kern-1mm-\kern-1mm\int_{\lambda
N_0(2^{r_0}\kern-0.5mm-\kern-0.5mm1)}^{+\infty}\int_{0}^{\frac{g_1Q_{pk}}{N_0\left(2^{r_0}\kern-0.5mm-\kern-0.5mm1\right)}}
f_0(g_0)f_1(g_1)dg_0dg_1,
\end{align}
where $\lambda$ is the nonnegative dual variable associated with
\eqref{con-transmit-av} in $\mathscr{F}_{3}$. If
\eqref{con-transmit-av} in $\mathscr{F}_{3}$ is satisfied with
strict inequality, $\lambda$ must be zero. Otherwise, $\lambda$ can
be obtained by substituting \eqref{outage-power-ap} into the
constraint ${\rm E}[P(g_0,g_1)]=P_{av}$,

Theorem 4 can be proved similarly as Theorem 3; the proof is thus
omitted here. Clearly, the power control policy given in
\eqref{outage-power-ap} is also 2D-TCI.


\subsection{Average transmit power constraint and average interference power constraint}\label{caatai}
In this case, $\mathscr{F}$ in \eqref{outage-probability} becomes
$\mathscr{F}_{4}$. The optimal solution of
\eqref{outage-probability} in this case is given by the following
theorem.

\emph{Theorem 5:} The optimal solution of \eqref{outage-probability}
subject to the power constraints given in $\mathscr{F}_{4}$ is
\begin{align}\label{outage-power-aa}
P(g_0,g_1)=\left\{
\begin{array}{ccc}
  \frac{N_0 \left(2^{r_0}-1\right)}{g_1},  & \lambda+\mu g_0<\frac{g_1}{N_0\left(2^{r_0}-1\right)}\\
  0,  & \mbox{otherwise}
\end{array}
\right.,
\end{align}
where $\lambda$ and $\mu$  are the nonnegative dual variables
associated with \eqref{con-transmit-av} and
\eqref{con-interference-av} in $\mathscr{F}_{4}$, respectively. If
\eqref{con-transmit-av} or \eqref{con-interference-av} in
$\mathscr{F}_{4}$ is satisfied with strict inequality, $\lambda$ or
$\mu$ must be zero correspondingly. Otherwise, $\lambda$ and $\mu$
can be jointly determined by substituting \eqref{outage-power-aa}
into the constraints ${\rm E}[P(g_0,g_1)]=P_{av}$ and ${\rm E}[g_0
P(g_0,g_1)]=Q_{av}$.


Theorem 5 can be proved similarly as Theorem 3.

\subsection{Analytical Results}
In this part, we provide the analytical results for the minimum
outage probability under only the peak or the average interference
power constraint.

\subsubsection{Peak interference power
constraint only}

From \eqref{outage-power-pp}, by setting $P_{pk}=+\infty$, we have
\begin{align}\label{eq:optimalPcase1}
P(g_0,g_1)=\frac{Q_{pk}}{g_0}.
\end{align}
Substituting (\ref{eq:optimalPcase1}) into
\eqref{outage-probability} yields
\begin{align}\label{eq:Pout-shortterm}
\mathscr{P}_{out}=Pr\left\{\frac{g_1}{g_0}< \frac{N_0
\left(2^{r_0}-1\right)}{Q_{pk}} \right\}.
\end{align}

In the following, the minimum outage probability is evaluated under
different fading models.

\emph{a) Rayleigh fading:}  Since $\frac{g_1}{g_0}$ has the same PDF
as $\frac{g_0}{g_1}$, with the PDF of $\frac{g_0}{g_1}$ given in
\eqref{pdf-rayleigh}, we have
\begin{align}
\mathscr{P}_{out}\kern-1mm=\kern-1mm\int_0^\frac{N_0
\left(2^{r_0}\kern-0.5mm-\kern-0.5mm1\right)}{Q_{pk}}\kern-0.5mm
\frac{1}{(x\kern-0.5mm+\kern-0.5mm1)^{2}} dx
\kern-0.5mm=\kern-0.5mm1\kern-1mm-\kern-1mm\frac{Q_{pk}}{N_0
\left(2^{r_0}\kern-0.5mm-\kern-0.5mm1\right)
\kern-0.5mm+\kern-0.5mmQ_{pk}}\label{eq:Pout-vs-r0}.
\end{align}


\emph{b) Nakagami fading:} With the PDF of $\frac{g_1}{g_0}$ given
in \eqref{pdf-nakagami} (note that $\frac{g_1}{g_0}$ has the same
PDF as $\frac{g_0}{g_1}$), we have
\begin{align}\label{eq:Pout-shortterm}
\mathscr{P}_{out}=\int_0^\frac{N_0 \left(2^{r_0}-1\right)}{Q_{pk}}
\frac{x^{m-1}}{\mathcal {B}(m,m)(x+1)^{2m}} dx=\frac{1}{\mathcal
{B}(m,m)}\int_0^\frac{N_0 \left(2^{r_0}-1\right)}{Q_{pk}}
\frac{x^{m-1}}{(x+1)^{2m}} dx.
\end{align}

From (3.194-1) in \cite{Tablesofintegral}, the above equation is
simplified as
\begin{align}
\mathscr{P}_{out}=\frac{1}{m\mathcal {B}(m,m)}\left[\frac{N_0
\left(2^{r_0}-1\right)}{Q_{pk}} \right]^m\left\{
{_2F_1}\left(2m,m;m+1;-\frac{N_0
\left(2^{r_0}-1\right)}{Q_{pk}}\right) \right\},
\end{align}
where ${_2F_1}(a, b ; c ; x)$ is known as Gauss's hypergeometric
function \cite{Tablesofintegral}.


\emph{c) Log-normal fading:} With the PDF of $\frac{g_1}{g_0}$ given
in Section \ref{section:delay-limited capacity} (note that
$\frac{g_1}{g_0}$ has the same PDF as $\frac{g_0}{g_1}$), we have
\begin{align}
\mathscr{P}_{out}=Pr\left\{e^Y< \frac{N_0
\left(2^{r_0}-1\right)}{Q_{pk}} \right\}=1-\frac{1}{2} {\rm erfc}
\left(\frac{1}{2\sigma} \log\left[\frac{N_0
\left(2^{r_0}-1\right)}{Q_{pk}} \right]\right),
\end{align}
where ${\rm erfc}(\cdot)$ is defined as ${\rm erfc}(t)\triangleq
\frac{2}{\sqrt{\pi}}\int_t^\infty e^{-x^2}dx $.
\subsubsection{Average interference power constraint only}
From \eqref{outage-power-pa}, by setting $P_{pk}=+\infty$ and
denoting $\omega^{\ast}=\frac{1}{\lambda N_0 (2^{r_0}-1)}$, we have
\begin{align}\label{eq:optimal-P-av} P(g_0,g_1)=\left\{\begin{array}{cc}
           \frac{N_0 \left(2^{r_0}-1\right)}{g_1}, & \frac{g_0}{g_1} <\omega^{\ast}\\
           0, & \mbox{otherwise}
         \end{array}
\right.,
\end{align}
and the minimum outage probability is given by
\begin{align}
\mathscr{P}_{out}=1-Pr\left\{ \frac{g_0}{g_1}<\omega^{\ast}\right\},
\end{align}
where $\omega^{\ast}$ is obtained by substituting
(\ref{eq:optimal-P-av}) into the constraint ${\rm E}[g_0
P(g_0,g_1)]= Q_{av}$.

In the following, the minimum outage probability is evaluated under
different fading models.

\emph{a) Rayleigh fading:} With the PDF of $\frac{g_0}{g_1}$ given
in \eqref{pdf-rayleigh}, we have
\begin{align}\label{eq:l-rayl-Pout}
\mathscr{P}_{out}&=1-\int_0^{\omega^{\ast}} \frac{1}{(x+1)^{2}}
dx=\frac{1}{1+\omega^{\ast}},
\end{align}
where $\omega^{\ast}$ is given by
\begin{align}\label{eq:32}
\int_0^{\omega^{\ast}}\frac{x}{(x+1)^{2}}dx=\frac{Q_{av}}{N_0
\left(2^{r_0}-1\right)}.
\end{align}
From (\ref{eq:32}), we have
\begin{align}\label{eq:33}
\omega^{\ast}\kern-0.5mm=\kern-0.5mm\exp&\Bigg[{\rm \mathcal {W}
\left(-e^{-\kern-0.5mm1\kern-0.5mm-\kern-0.5mm\frac{Q_{av}}{N_0
\left(2^{r_0}\kern-0.5mm-\kern-0.5mm1\right)}}\right)}
\kern-0.5mm+\kern-0.5mm1\kern-0.5mm+\kern-0.5mm\frac{Q_{av}}{N_0
\left(2^{r_0}\kern-0.5mm-\kern-0.5mm1\right)}\Bigg]\kern-1mm-\kern-1mm1,
\end{align}
where $\mathcal {W} (x)$ is the Lambert-W function, which is defined
as the inverse function of $f(w)=we^w$.

As can be seen from (\ref{eq:l-rayl-Pout}), if $\omega^{\ast}$ goes
to infinity, the outage probability becomes zero; however, from
(\ref{eq:33}), it is seen that $\omega^{\ast}$ is infinity only when
$r_0=0$. This indicates that the zero-outage capacity for Rayleigh
fading is zero, which is consistent with the result obtained in
Section \ref{section:delay-limited capacity}.

\emph{b) Nakagami fading:} With the PDF of $\frac{g_0}{g_1}$ given
in \eqref{pdf-nakagami}, we have
\begin{align}\label{eq:Pout-shortterm}
\mathscr{P}_{out}=1-\int_0^{\omega^{\ast}} \frac{x^{m-1}}{\mathcal
{B}(m,m)(x+1)^{2m}} dx
=1-\frac{\left(\omega^{\ast}\right)^m}{m\mathcal {B}(m,m)} {_2F_1}
\left(2m,m;m+1;-\omega^{\ast}\right),
\end{align}
where $\omega^{\ast}$ is given by
\begin{align}
\frac{1}{\mathcal
{B}(m,m)}\int_{0}^{\omega^{\ast}}\frac{x^{m}}{(x+1)^{2m}}dx=\frac{Q_{av}}{N_0
\left(2^{r_0}-1\right)}.
\end{align}
From (3.194-1) in \cite{Tablesofintegral},  the above equation is
simplified as
\begin{align}\label{eq:37}
\frac{\left(w^{\ast}\right)^{m+1} {_2F_1}
\left(2m,m+1;m+2;-w^{\ast}\right)}{(m+1)\mathcal {B}(m,m)}
=\frac{Q_{av}}{N_0 \left(2^{r_0}-1\right)}.
\end{align}

From the above, for the case of $m=2$, the outage probability can be
shown to be $
\mathscr{P}_{out}=\frac{1+3\omega^{\ast}}{\left(1+\omega^{\ast}\right)^3}
$, and $\omega^{\ast}$ satisfies $
2\left[1-\frac{1+3\omega^{\ast}+3\left(\omega^{\ast}\right)^2}{\left(1+\omega^{\ast}\right)^3}\right]=\frac{Q_{av}}{N_0
\left(2^{r_0}-1\right)}$. From the above two formulas, when
$\omega^{\ast}$ is infinity, the outage probability becomes zero and
$r_0$ becomes the delay-limited capacity
$\log_{2}\left(1+\frac{Q_{av}}{2N_0 }\right)$. This is consistent
with the result obtained in Section \ref{section:delay-limited
capacity}.

\emph{c) Log-normal fading:}  With the PDF of $\frac{g_0}{g_1}$
given in Section \ref{section:delay-limited capacity}, we have
\begin{align}\label{eq:l-lognormal-Pout}
\mathscr{P}_{out}&=1-Pr\left\{e^Y<\omega^{\ast} \right\}
=\frac{1}{2} {\rm erfc} \left(\frac{1}{2\sigma}
\log\left(\omega^{\ast} \right)\right),
\end{align}
where $\omega^{\ast}$ is determined by
\begin{align}\label{eq:41}
\int_{-\infty}^{\log\left(\omega^{\ast}\right)}\kern-5pte^y
\frac{1}{\sqrt{2\pi}\left(\sqrt{2}\sigma\right)}
\exp{\left(-\frac{y^2}{2\times
2\sigma^2}\right)}dy=\frac{Q_{av}}{N_0 \left(2^{r_0}-1\right)}.
\end{align}
The above equation can be simplified to
\begin{align}\label{eq:66}
e^{\sigma^2}\left[1-\frac{1}{2} {\rm erfc}
\left(\frac{\log\left(\omega^{\ast} \right)-2\sigma^2}{2\sigma}
\right) \right]=\frac{Q_{av}}{N_0 \left(2^{r_0}-1\right)}.
\end{align}

It is seen from (\ref{eq:l-lognormal-Pout}), the zero-outage
probability is achieved when $\omega^{\ast}$ goes to infinity. It is
clear from (\ref{eq:66}) that, when $\omega^{\ast}$ goes to
infinity, $ r_0=\log_{2}\left(1+\frac{Q_{av}}{N_0
e^{\sigma^2}}\right)$. Again, this is consistent with the
delay-limited capacity obtained in Section
\ref{section:delay-limited capacity}.

\section{Simulation Results}\label{NumericalResults}
In this section, we present and discuss the simulation results for
the capacities of the SU fading channels under spectrum sharing with
the proposed power allocation strategies.

\subsection{Ergodic capacity}
In this subsection, the simulation results for ergodic capacity are
presented. For Rayleigh fading channels, the channel power gains
(exponentially distributed) are assumed to be unit mean. For AWGN
channels, the channel power gains are also assumed to be one.

\begin{figure}[t]
        \centering
        \includegraphics*[width=12cm]{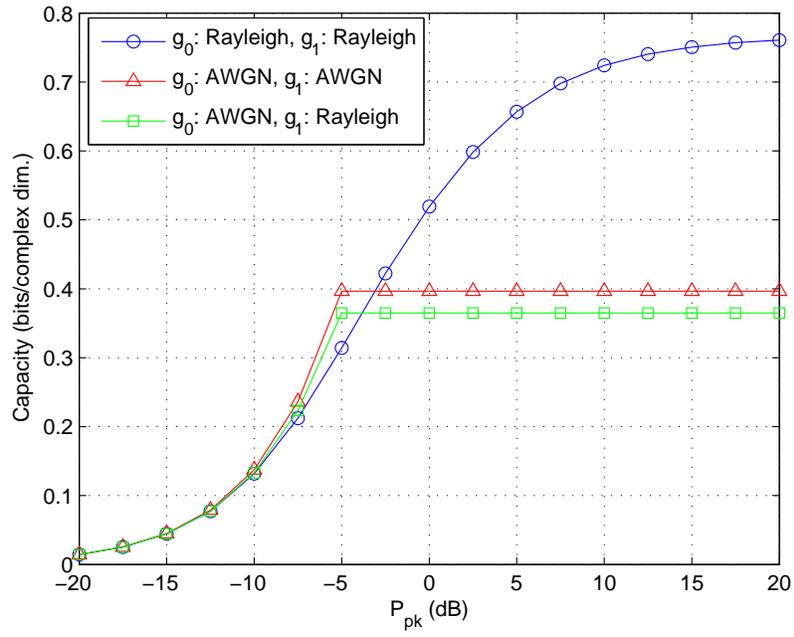}
        \caption{Ergodic capacity vs. $P_{pk}$ with $Q_{pk}=-5dB$ for different channel models.}
        \label{case1}
\end{figure}

Fig. \ref{case1} shows the ergodic capacity under peak transmit and
peak interference power constraints for $Q_{pk}=-5dB$. It is
observed that when $P_{pk}$ is very small, the ergodic capacities
for the three curves shown in this figure are almost the same. This
indicates that $P_{pk}$ limits the performance of the network.
However, when $P_{pk}$ is sufficiently large compared with $Q_{pk}$,
the ergodic capacities become different. In this case, when $g_0$
models the AWGN channel, the capacity of SU link when $g_1$ also
models the AWGN channel is higher than that when $g_1$ models the
Rayleigh fading channel. This indicates that fading of the SU
channel is harmful. However, when $g_1$ models the Rayleigh fading
channel, the capacity for SU link when $g_0$ models the AWGN channel
is lower than that when $g_0$ models the Rayleigh fading channel.
This illustrates that fading of the channel between SU-Tx and PU-Rx
is a beneficial factor in terms of maximizing the ergodic capacity
of SU channel.

\begin{figure}[t]
        \centering
        \includegraphics*[width=12cm]{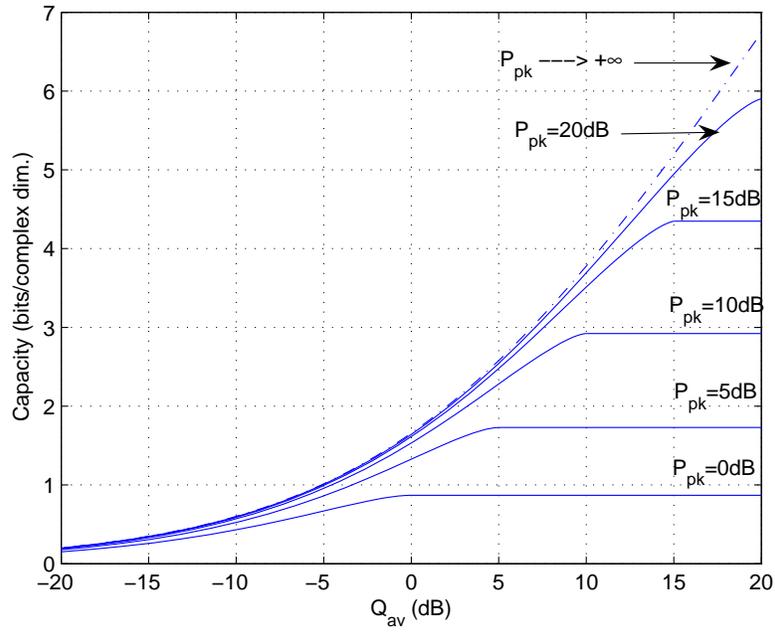}
        \caption{Ergodic capacity under peak transmit and average interference power constraints.}
        \label{case2}
\end{figure}

Fig. \ref{case2} shows the ergodic capacity versus $Q_{av}$ under
peak transmit and average interference power constraints. For
comparison, the curve with $P_{pk}=+\infty$ (i.e. no transmit power
constraint) is also shown. It is observed that when $Q_{av}$ is
small, the capacities for different $P_{pk}$'s do not vary much.
This illustrates that $Q_{av}$ limits the achievable rate of SU.
However, when $P_{pk}$ is sufficiently large compared to $Q_{av}$,
the capacities become flat. This indicates that $P_{pk}$ becomes the
dominant constraint in this case. Furthermore, with $P_{pk}$ being
sufficiently large, the ergodic capacity of SU channel becomes close
to that without transmit power constraint.

\begin{figure}[t]
        \centering
        \includegraphics*[width=12cm]{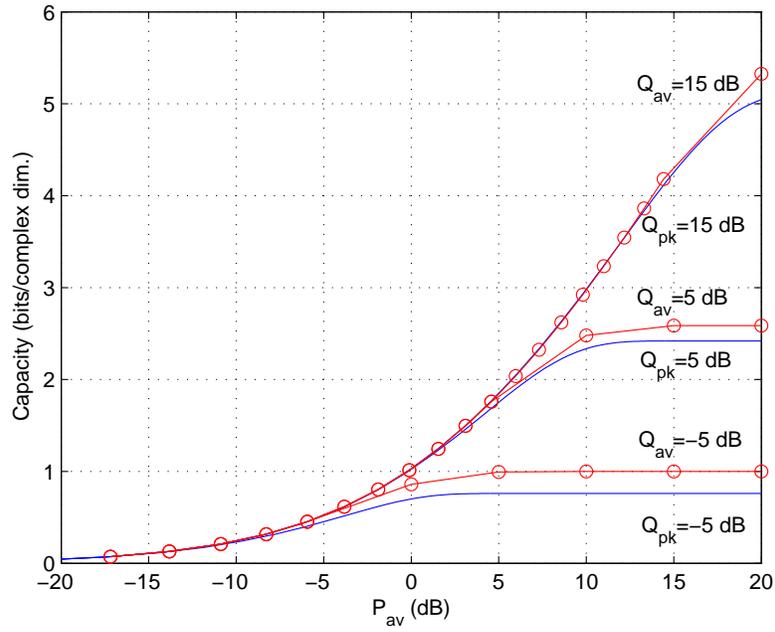}
        \caption{Ergodic capacity vs. $P_{av}$ under peak or average interference power constraints.}
        \label{case3}
\end{figure}

Fig. \ref{case3} shows the ergodic capacity versus  $P_{av}$ under
different types of interference power constraints. As shown in the
figure, the ergodic capacity under average interference power
constraint is larger than that under peak interference power
constraint with the same value of $P_{av}$. This is because the
power control of SU is more flexible under average over peak
interference power constraint.

\subsection{Delay-limited capacity and outage capacity}
In this subsection, the simulation results for delay-limited and
outage capacities are presented. For Rayleigh fading channels, the
channel power gains (exponentially distributed) are assumed to be
unit mean. Besides, $m=2$ is chosen for the unit-mean Nakagami
fading channels used in the simulation. For log-normal fading
channels, $\sigma^2=1$ is used. This is because log-normal shadowing
is usually characterized in terms of its dB-spread $\sigma_{dB}$,
which ranges from $4 dB$ to $12 dB$ by empirical measurements, and
is related to $\sigma$ by $\sigma=0.1\log(10)\sigma_{dB}$
\cite{Ghasemi}. We thus choose $\sigma^2=1$ as this value of
$\sigma$ makes the dB-spread lying within its typical ranges.

\begin{figure}[t]
\centering
\includegraphics*[width=12cm]{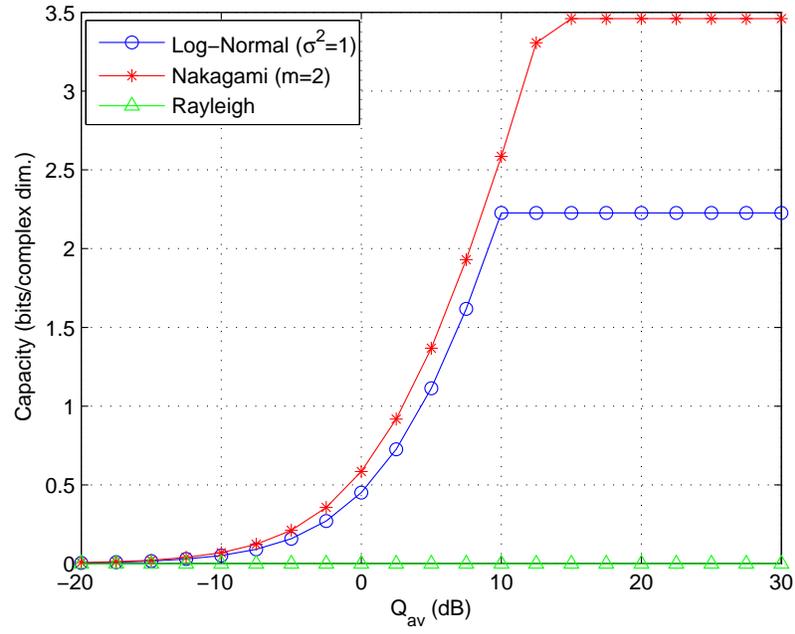}
\caption{Delay-limited capacity  vs. $Q_{av}$ with $P_{av}=10dB$ for
different fading channel models. } \label{delaycase}
\end{figure}

Fig. \ref{delaycase} shows the delay-limited capacity under
$P_{av}=10dB$ for different fading models versus $Q_{av}$. It is
seen that the delay-limited capacity for Nakagami fading and
log-normal shadowing increases with $Q_{av}$. However, when $Q_{av}$
is sufficiently large, the delay-limited capacity will get saturated
due to $P_{av}$. Note that the delay-limited capacity of Rayleigh
fading model is zero regardless of $Q_{av}$. This is consistent with
our analysis in Section \ref{section:delay-limited capacity}.

\begin{figure}[t]
        \centering
        \includegraphics*[width=12cm]{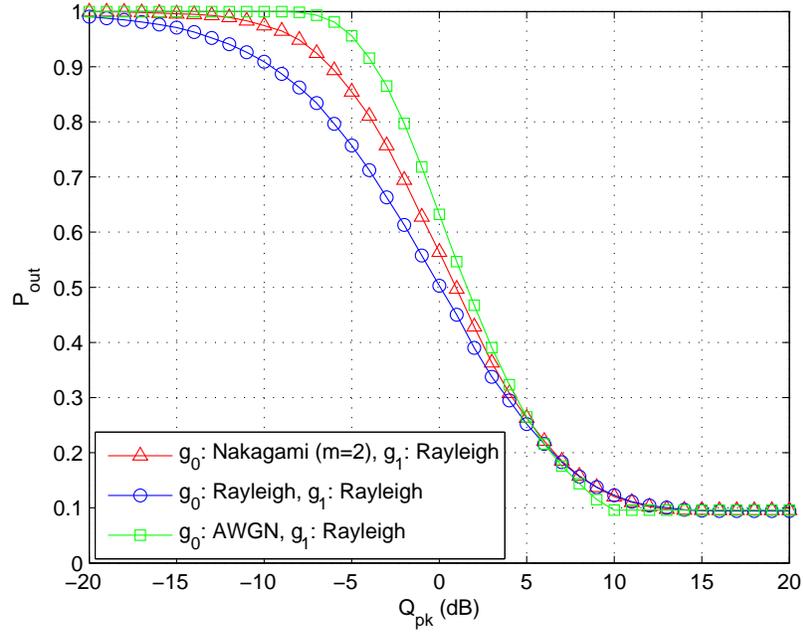}
        \caption{Outage probability  vs. $Q_{pk}$ for $r_0=1$ bit/complex dim. $P_{pk}=10dB$ for different fading channel models.}
        \label{outcase1}
\end{figure}

Fig. \ref{outcase1} shows the outage probability for different
fading models under $P_{pk}=10dB$ and $r_0=1$ bit/complex dimension
(dim.). It is seen that when $Q_{pk}$ is small, the outage
probability of SU link when $g_0$ models a fading channel is smaller
than that when $g_0$ models the AWGN channel. Besides, more severe
the fading is, the smaller the outage probability is. This
illustrates that fading of the channel between SU-Tx and PU-Rx is
good in terms of minimizing the outage probability of SU channel.
However, when $Q_{pk}$ has the same value of $P_{pk}$, the outage
probability when $g_0$ models a fading channel is larger than that
when $g_0$ models the AWGN channel. This can be foreseen from
\eqref{outage-power-pp}. When $Q_{pk}=P_{pk}$,  the channel
inversion condition for the AWGN case is $\frac{2^{r_0}-1}{g_1}\le
P_{pk}$. However, the channel inversion condition for the fading
case is $\frac{2^{r_0}-1}{g_1}\le \min(P_{pk}, \frac{Q_{pk}}{g_0})$,
which can be more restrictive than that in the AWGN case if $g_0>1$.
The higher the probability $g_0>1$ is, the larger the resultant
outage probability is. However, when $Q_{pk}$ is sufficiently large,
both fading and AWGN channels will have the same outage probability,
since $P_{pk}$ becomes the dominant constraint in this case.

\begin{figure}[t]
        \centering
        \includegraphics*[width=12cm]{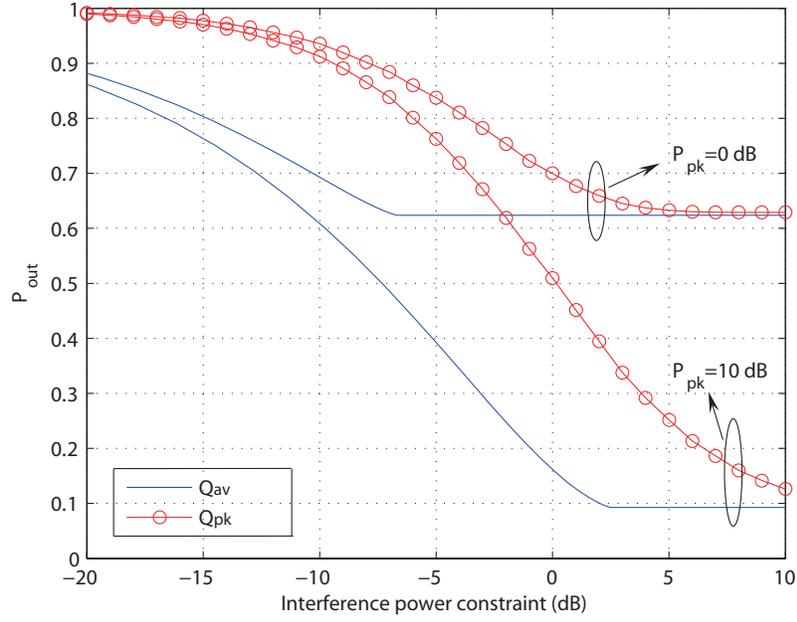}
        \caption{Outage probability for $r_0=1$ bit/complex dim. under peak or average interference power constraints}
        \label{outcase2}
\end{figure}

Fig. \ref{outcase2} shows the outage probability under peak and
average interference power constraints for $r_0=1$ bit/complex dim.
under $P_{pk}=0dB$ or $P_{pk}=10dB$. It is seen that under the same
$P_{pk}$, the outage probability under the average interference
power constraint is smaller than that under the peak interference
power constraint. This is due to the fact that the power control
policy of SU is more flexible under the average over the peak
interference power constraint.

\begin{figure}[t]
\centering
\includegraphics*[width=12cm]{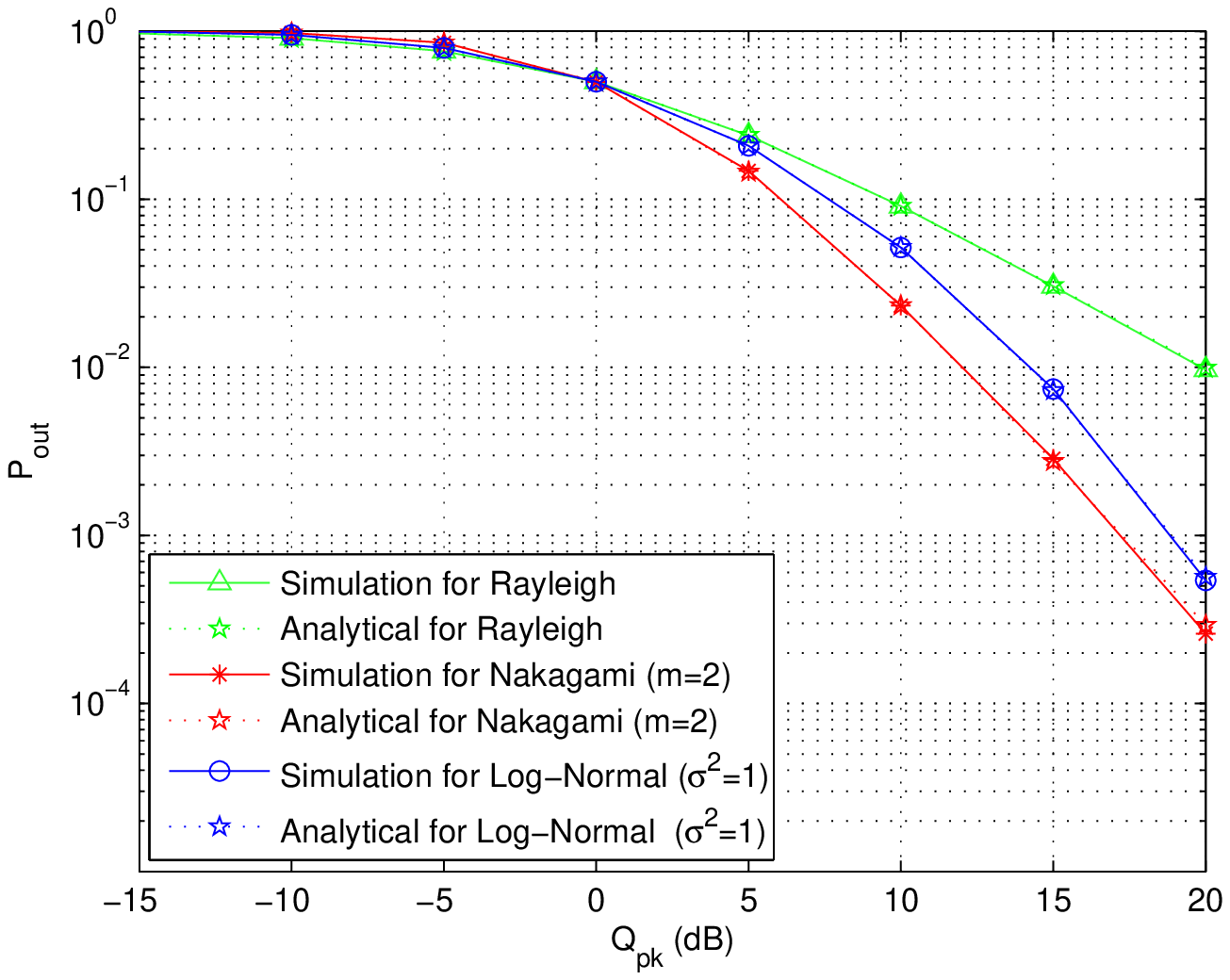}
\caption{Outage probability for $r_0=1$ bit/complex dim. under peak
interference power constraint only.} \label{soutcase1}
\end{figure}

Fig. \ref{soutcase1} shows the outage probability for different
fading models under the peak interference power constraint only with
$r_0=1$ bit/complex dim.. It is observed that the simulation results
match the analytical results very well. Moreover, it is observed
that the outage probability curves overlap when $Q_{pk}$ is very
small, indicating that the fading models do not affect the outage
probability notably for small value of $Q_{pk}$.

\begin{figure}[t]
\centering
\includegraphics*[width=12cm]{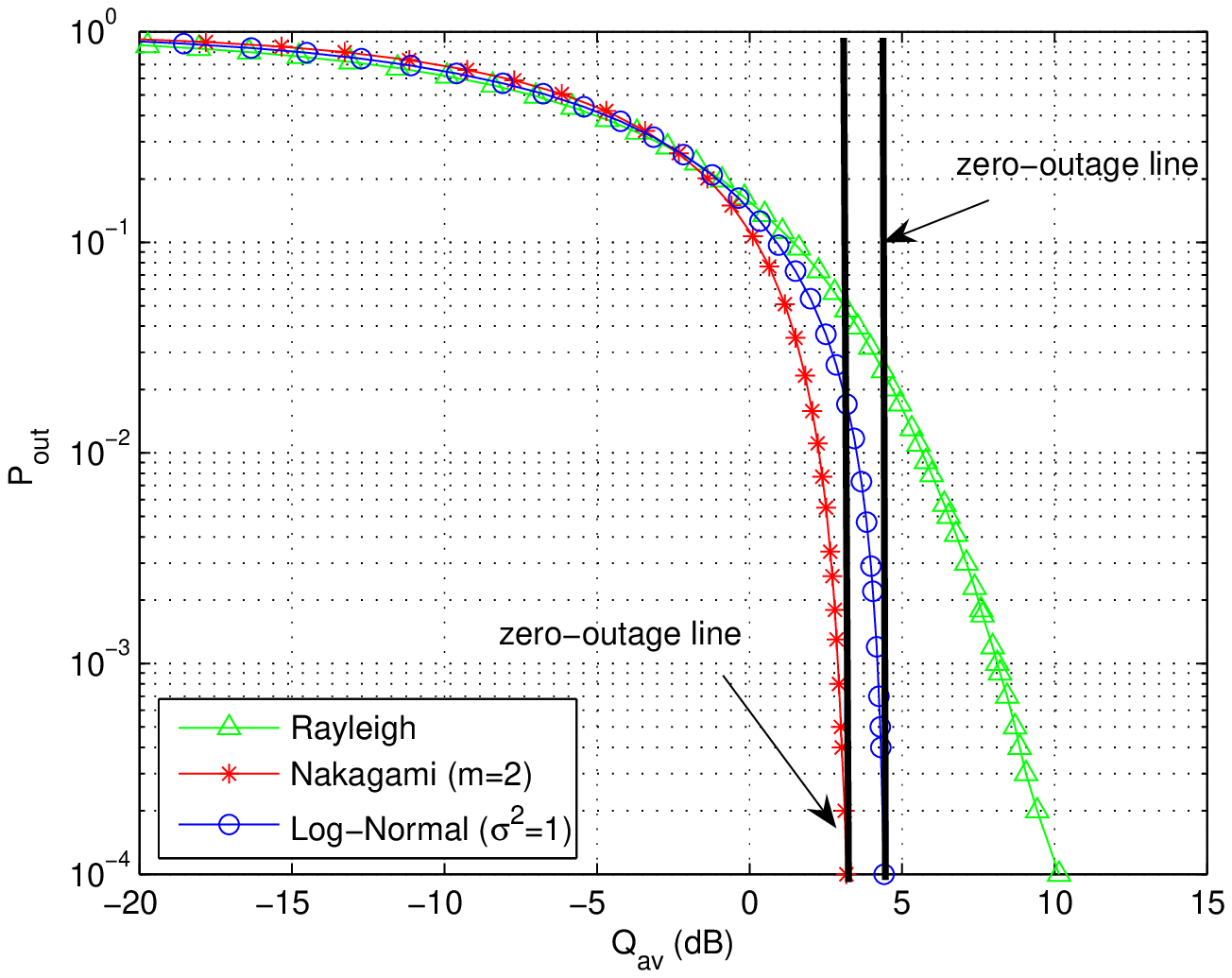}
\caption{Outage probability for $r_0=1$ bit/complex dim. under
average interference power constraint only.} \label{soutcase2}
\end{figure}

Fig. \ref{soutcase2} illustrates the outage capacity versus average
interference power constraint $Q_{av}$ when the target rate $r_0$ is
1 bit/complex dim.. It is observed that the outage probability for
Nakagami fading and log-normal shadowing drop sharply when $Q_{av}$
reaches a certain value. This demonstrates that when $Q_{av}$
approaches infinity, the outage probability becomes zero. In
contrast, there is no such an evident threshold observed for
Rayleigh fading channel, since its delay-limited capacity is zero.
Additionally, comparing Fig.s \ref{soutcase1} and \ref{soutcase2},
it is observed that the outage probability under average
interference power constraint is smaller than that under peak
interference power constraint when $Q_{av}=Q_{pk}$, suggesting that
the power allocation scheme under the former is more flexible over
the latter. Furthermore, comparing Fig. \ref{soutcase2} with Fig.
\ref{delaycase}, it is observed that $Q_{av}$ required to achieve
the zero-outage probability for $r_0=1$ bit/complex dim. is
consistent with that required to achieve the same delay-limited
capacity.

\section{Conclusions}\label{conclusion}
In this paper, the optimal power allocation strategies to achieve
the ergodic, delay-limited, and outage capacities of a SU fading
channel under spectrum sharing are studied, subject to different
combinations of peak/average transmit and/or peak/average
interference power constraints. It is shown that under the same
threshold value, average interference power constraints are more
flexible over their peak constraint counterparts to maximize SU
fading channel capacities. The effects of different fading channel
statistics on achievable SU capacities are also analyzed. One
important observation made in this paper is that fading of the
channel between SU-Tx and PU-Rx can be a good phenomenon for
maximizing the capacity of SU fading channel.

\section*{Appendix A\\ Proof of Theorem 1}
By introducing the dual variable associated with the average
interference power constraint, the partial Lagrangian of this
problem is expressed as
\begin{align}\label{Lagrangianforcase2}
L(P(g_0,g_1),\lambda)= {\rm E}\kern-1mm\left[\log_2
\left(\kern-0.5mm1\kern-0.5mm+\kern-0.5mm\frac{g_1P(g_0,g_1)}{N_0}\right)\kern-0.5mm\right]-\kern-1mm\lambda\left({\rm
E}[g_0P(g_0,g_1)]\kern-0.5mm-\kern-0.5mmQ_{av}\right)\kern-0.5mm,
\end{align}
where $\lambda$ is the nonnegative dual variable associated with the
constraint ${\rm E}[g_0P(g_0,g_1)]\le Q_{av}$.

Let $\mathcal {A}$ denote the set of $\left\{0\le P(g_0,g_1)\le
P_{pk}\right\}$. The dual function is then expressed as
\begin{align}
q(\lambda)=\max_{P(g_0,g_1) \in \mathcal{A}} L(P(g_0,g_1),\lambda).
\end{align}

The Lagrange dual problem is then defined as $\min_{\lambda \ge0}
q(\lambda)$. It can be verified that the duality gap is zero for the
convex optimization problem addressed here, and thus solving its
dual problem is equivalent to solving the original problem.
Therefore, according to the Karush-Kuhn-Tucker (KKT) conditions
\cite{Convexoptimization}, the optimal solutions needs to satisfy
the following equations:
\begin{eqnarray}\label{neqcase2}
0 \le P(g_0,g_1)\le P_{pk}, \quad {\rm
E}[g_0P(g_0,g_1)]\le Q_{av},\\
\label{gpeqqcase2}\lambda({\rm E}[g_0P(g_0,g_1)]-Q_{av})=0.
\end{eqnarray}

For a fixed $\lambda$, by dual decomposition \cite{Zhang08MAC}, the
dual function can be decomposed into a series of similar
sub-dual-functions each for one fading state. For a particular
fading state, the problem can be shown equivalent to
\begin{align}
\max_{P(g_0,g_1)}&\log_2
\left(\kern-0.5mm1\kern-0.5mm+\kern-0.5mm\frac{g_1P(g_0,g_1)}{N_0}\right)-\lambda
g_0P(g_0,g_1), \\
\mbox{s.t.}~~ &P(g_0,g_1)\le P_{pk},\label{appdix-53}\\
            &P(g_0,g_1)\ge 0\label{appdix-54}.
\end{align}

The dual function of this sub-problem is
\begin{align}
L_{sub}(P(g_0,g_1),\mu,\nu)\kern-1mm=\kern-1mm\log_2
\left(\kern-0.5mm1\kern-0.5mm+\kern-0.5mm\frac{g_1P(g_0,g_1)}{N_0}\right)\kern-1mm-\kern-1mm\lambda
g_0P(g_0,g_1)\kern-1mm-\kern-1mm\mu(P(g_0,g_1)\kern-1mm-\kern-1mm
P_{pk})+\nu P(g_0,g_1),
\end{align}
where $\mu$ and $\nu$ are the nonnegative dual variables associated
with the constraints \eqref{appdix-53} and \eqref{appdix-54},
respectively.

The sub-dual problem is then defined as
$q_{sub}(\mu,\nu)=\min_{\mu\ge0,\nu\ge0}
L_{sub}(P(g_0,g_1),\mu,\nu).$ This is also a convex optimization
problem for which the duality gap is zero. Therefore, according to
the KKT conditions, the optimal solutions needs to satisfy the
following equations:
\begin{eqnarray}
\label{peqpcase2}\mu(P(g_0,g_1)- P_{pk})=0,\\
\label{nupcase2}\nu P(g_0,g_1)=0,\\
\label{dLagcase2}\frac{Kg_1}{g_1P(g_0,g_1)+N_0}-\lambda
g_0-\mu+\nu=0.
\end{eqnarray}
From (\ref{dLagcase2}), it follows
\begin{eqnarray}\label{tpcase2}
P(g_0,g_1)=\frac{K}{\mu-\nu+\lambda g_0}-\frac{N_0}{g_1}.
\end{eqnarray}

Suppose that $P(g_0,g_1)<P_{pk}$, when $g_0\le
\frac{K}{\lambda(P_{pk}+\frac{N_0}{g_1})}$ or equivalently
$(\frac{K}{\lambda g_0}-\frac{N_0}{g_1})\ge P_{pk}$. Then, from
(\ref{peqpcase2}), it follows that $\mu=0$. Therefore,
(\ref{tpcase2}) reduces to $P(g_0,g_1)=\frac{K}{-\nu +\lambda
g_0}-\frac{N_0}{g_1}$. Then $ P(g_0,g_1)<P_{pk}$ results in
$\frac{K}{-\nu +\lambda g_0}-\frac{N_0}{g_1}<P_{pk}$. Since $\nu
\ge0$, it follows that $P_{pk}>\frac{K}{-\nu +\lambda
g_0}-\frac{N_0}{g_1}\ge \frac{K}{\lambda g_0}-\frac{N_0}{g_1}$. This
contradicts the presumption. Therefore, from \eqref{neqcase2}, it
follows that
\begin{align} \label{eq:ergodiccase2-proof1}
P(g_0,g_1)=P_{pk}, \quad\mbox{if}\quad g_0\le
\frac{K}{\lambda(P_{pk}+\frac{N_0}{g_1})}.
\end{align}

Suppose $P(g_0,g_1)>0$, when $g_0\ge \frac{Kg_1}{\lambda N_0}$ or
equivalently $\frac{K}{\lambda g_0}-\frac{N_0}{g_1}\le 0$. Then,
from \eqref{nupcase2}, it follows that $\nu=0$. Therefore,
(\ref{tpcase2}) reduces to $P(g_0,g_1)=\frac{K}{\mu +\lambda
g_0}-\frac{N_0}{g_1}$. Then $ P(g_0,g_1)>0$ results in
$\frac{K}{\mu+\lambda g_0}-\frac{N_0}{g_1}>0$. Since $\mu \ge 0$, it
follows that $\frac{K}{\lambda g_0}-\frac{N_0}{g_1}\ge \frac{K}{\mu
+\lambda g_0}-\frac{N_0}{g_1}>0 $. This contradicts with the
presumption. Therefore, from \eqref{neqcase2}, it follows
\begin{align}\label{eq:ergodiccase2-proof2}
P(g_0,g_1)=0, \quad\mbox{if}\quad g_0\ge \frac{Kg_1}{\lambda N_0}.
\end{align}

Suppose $P(g_0,g_1)=0$, when $\frac{Kg_1}{\lambda
N_0}>g_0>\frac{K}{\lambda(P_{pk}+\frac{N_0}{g_1})}$ or equivalently
$0<\frac{K}{\lambda g_0}-\frac{N_0}{g_1}<P_{pk}$. Then, from
(\ref{peqpcase2}), it follows that $\mu=0$. Therefore,
(\ref{tpcase2}) reduces to $P(g_0,g_1)=\frac{K}{-\nu +\lambda
g_0}-\frac{N_0}{g_1}$. Then $ P(g_0,g_1)=0$ results in
$\frac{K}{-\nu +\lambda g_0}-\frac{N_0}{g_1}=0$. Since $\nu \ge0$,
it follows that $0>\frac{K}{-\nu +\lambda g_0}-\frac{N_0}{g_1}\ge
\frac{K}{\lambda g_0}-\frac{N_0}{g_1}$. This contradicts the
presumption. Therefore, $P(g_0,g_1)\neq 0$ for this set of $g_0$.
Next, suppose $P(g_0,g_1)=P_{pk}$ for the same set of $g_0$. Then,
from \eqref{nupcase2}, it follows that $\nu=0$. Therefore,
(\ref{tpcase2}) reduces to $P(g_0,g_1)=\frac{K}{\mu +\lambda
g_0}-\frac{N_0}{g_1}$. Then $P(g_0,g_1)=P_{pk}$ indicates
$\frac{K}{\mu +\lambda g_0}-\frac{N_0}{g_1}=P_{pk}$. Since $\mu \ge
0$, it follows $\frac{K}{\lambda g_0}-\frac{N_0}{g_1}\ge
\frac{K}{\mu +\lambda g_0}-\frac{N_0}{g_1}=P_{pk}$. This contradicts
the presumption. Therefore, $P(g_0,g_1)\neq P_{pk}$ for this set of
$g_0$. Now, from \eqref{nupcase2}, $P(g_0,g_1) \neq 0$ results in
$\nu=0$. From (\ref{peqpcase2}), $P(g_0,g_1) \neq P_{pk}$ results in
$\mu=0$. Therefore, from (\ref{tpcase2}), it follows
\begin{align}\label{eq:ergodiccase2-proof3}
P(g_0,g_1)=\frac{K}{\lambda g_0}-\frac{N_0}{g_1},
\quad\mbox{if}\quad \frac{Kg_1}{\lambda
N_0}>g_0>\frac{K}{\lambda(P_{pk}+\frac{N_0}{g_1})}.
\end{align}

From \eqref{gpeqqcase2}, it is easy to observe that $\lambda$ is
either equal to zero or determined by solving ${\rm
E}[g_0P(g_0,g_1)]=Q_{av}$.

Theorem 1 is thus proved.

\section*{Appendix B\\ Proof of Theorem 3}
The proof is organized in two steps. First, we show that the
solution of \eqref{outage-probability} subject to $\mathscr{F}_{2}$
must have the same structure as \eqref{outage-power-pa}. Secondly,
we show that $\lambda$ is determined by substituting
\eqref{outage-power-pa} into the constraint  ${\rm E}\left[g_0
P(g_0,g_1)\right]=Q_{av}$.

\emph{Step 1:} Define an indicator function,
\begin{align}
\chi=\left\{ \begin{array}{cc}
              1, & \log_{2}\left(1+\frac{g_1
P\left(g_0,g_1\right)}{N_0}\right)< r_0\\
              0, & \mbox{otherwise}
             \end{array}
\right..
\end{align}

Then the optimization problem \eqref{outage-probability} subject to
$\mathscr{F}_{2}$ can be rewritten as
\begin{align}
\min_{P(g_0,g_1)\in\mathscr{F}_{2}}{\rm E}\left\{\chi\right\}.
\end{align}

By introducing the dual variable $\lambda$ associated with the
average interference power constraint, the partial Lagrangian of
this problem is expressed as
\begin{align}
L\left(P(g_0,g_1),\lambda\right)={\rm E}\left\{\chi\right\}+\lambda
\left({\rm E}\{g_0P(g_0,g_1)\}-Q_{av}\right).
\end{align}

Let $\mathcal {A}$ denote the set of $\left\{P(g_0,g_1): 0\le
P(g_0,g_1)\le P_{pk}\right\}$. The dual function is then expressed
as
\begin{align}
\min_{P(g_0,g_1)\in \mathcal {A}}{\rm E}\left\{\chi\right\}+\lambda
\left({\rm E}\{g_0P(g_0,g_1)\}-Q_{av}\right).
\end{align}

For a fixed $\lambda$, by dual decomposition, the dual function can
be decomposed into a series of similar sub-dual-functions each for
one fading state. For a particular fading state, the problem can be
shown equivalent to
\begin{align}\label{ProofB-obj}
\min_{P(g_0,g_1)}~&\chi+\lambda g_0P(g_0,g_1),\\
\mbox{s.t.}~~~&P(g_0,g_1)\le P_{pk},\\
&P(g_0,g_1)\ge 0.
\end{align}

When $\chi=1$, \eqref{ProofB-obj} is minimized if $P(g_0,g_1)=0$,
and the minimum value of \eqref{ProofB-obj} is $1$; when $\chi=0$,
\eqref{ProofB-obj} is minimized if
$P(g_0,g_1)=\frac{N_0\left(2^{r_0}-1\right)}{g_1}$, and the minimum
value of \eqref{ProofB-obj} is $\lambda g_0
\frac{N_0\left(2^{r_0}-1\right)}{g_1}$. Thus,
$P(g_0,g_1)=\frac{N_0\left(2^{r_0}-1\right)}{g_1}$ is the optimal
solution of the problem, only when $\lambda g_0
\frac{N_0\left(2^{r_0}-1\right)}{g_1}<1$ and
$\frac{N_0\left(2^{r_0}-1\right)}{g_1} \le P_{pk}$ are satisfied
simultaneously. Otherwise, $P(g_0,g_1)=0$ is the optimal solution of
the problem. Therefore, the optimal solution has the same structure
as \eqref{outage-power-pa}.

\emph{Step 2:} Suppose $P^\ast(g_0,g_1)$ is the optimal solution of
\eqref{outage-probability} subject to $\mathscr{F}_{2}$ with
$\lambda=\lambda^\ast>0$ satisfying ${\rm E}\left[g_0
P^\ast(g_0,g_1)\right]<Q_{av}$. Suppose $P^\prime(g_0,g_1)$ is a
solution of \eqref{outage-probability} subject to $\mathscr{F}_{2}$
with $\lambda=\lambda^\prime>0$, which satisfies ${\rm E}\left[g_0
P^\prime(g_0,g_1)\right]=Q_{av}$. Then, it is easy to verify that
$\lambda^\ast>\lambda^\prime$. Therefore, from \eqref{26}, it
follows
\begin{align}
\mathscr{P}^\ast_{out}>\mathscr{P}^\prime_{out}
\end{align}
where the inequality results from the fact that
$\lambda^\ast>\lambda^\prime$ and $\mathscr{P}_{out}$ is an
increasing function with respect to $\lambda$. This result
contradicts our presumption. Therefore, the optimal $\lambda$ must
be determined by solving ${\rm E}\left[g_0
P(g_0,g_1)\right]=Q_{av}$. Otherwise,  if $\lambda=0$, the power
allocation strategy obtained in step 1 reduces to the truncated
channel inversion given in \cite{Goldsmith}, and this holds only
when ${\rm E}\left[g_0 P(g_0,g_1)\right]<Q_{av}$.

Theorem 3 is thus proved.

\bibliographystyle{IEEEtran}
\bibliography{CR}

\newpage

\end{document}